\documentclass[reprint,amsmath,amssymb,aps,prl,floatfix]{revtex4-2}
\usepackage[english]{babel}

\usepackage{graphicx}
\usepackage{dcolumn}
\usepackage{bm}
\usepackage{hyperref}
\usepackage{placeins}
\usepackage{needspace}
\hypersetup{hidelinks}

\begin{document}

\title{Nanohertz Loss in a Pendulum for\\
Macroscopic Entanglement under Structural Damping}

\author{Azusa Sawada}
\affiliation{Department of Physics, Faculty of Science, Gakushuin University, 1-5-1 Mejiro, Toshima, Tokyo 171-8588, Japan}

\author{Hina Nakano}
\affiliation{Department of Physics, Faculty of Science, Gakushuin University, 1-5-1 Mejiro, Toshima, Tokyo 171-8588, Japan}

\author{Kanta Watanabe}
\affiliation{Department of Physics, Faculty of Science, Gakushuin University, 1-5-1 Mejiro, Toshima, Tokyo 171-8588, Japan}

\author{Gaku Ohashi}
\affiliation{Department of Physics, Faculty of Science, Gakushuin University, 1-5-1 Mejiro, Toshima, Tokyo 171-8588, Japan}

\author{Shota Okumura}
\affiliation{Department of Physics, Faculty of Science, Gakushuin University, 1-5-1 Mejiro, Toshima, Tokyo 171-8588, Japan}

\author{Nobuyuki Matsumoto}
\thanks{Corresponding author: \texttt{matsumoto.granite@gmail.com}}
\affiliation{Department of Physics, Faculty of Science, Gakushuin University, 1-5-1 Mejiro, Toshima, Tokyo 171-8588, Japan}

\begin{abstract}
Pendulums are attractive for macroscopic quantum control because gravitational
dissipation dilution lowers the modal loss, while the $1/f$ force noise of
structural damping allows nearly lossless trapping to suppress the noise at an
upshifted resonance. The same $1/f$ spectrum, however, produces a low-frequency
tail that penalizes entanglement. To overcome this penalty, we realize a $7$-mg
pendulum suspended by a stepped fused-silica fiber. Ring-down measurements give
an energy-decay rate $\Gamma/2\pi=361(39)$ nHz at the bare resonance
$\omega_0/2\pi=2.63$ Hz.
The reduction in $\omega_0\Gamma$ relative to the previous uniform suspension
more than offsets the penalty. Although both suspensions support entanglement
in the error-free model, only the stepped one allows robust verification in
the multimode model with experimental uncertainties.
\end{abstract}

\maketitle
\raggedbottom

\textit{Introduction}.---Quantum entanglement is a central resource for quantum information processing~\cite{Horodecki2009} and can enable quantum-enhanced sensing~\cite{Pezze2018}. At macroscopic scales, entanglement generated by gravity between spatially separated masses can serve as a witness to the quantum nature of gravity~\cite{PhysRevLett.119.240401,PhysRevLett.119.240402}.
Continuous measurement can conditionally entangle suspended mirrors from milligram to kilogram scales~\cite{PhysRevA.107.032410,PhysRevLett.100.013601}, enabling tests of semiclassical gravity models and gravity-induced entanglement~\cite{Yang2013SN,Liu2023CCSN,Miki2024GIE,Miki2025CCSN}. 
Reaching this regime requires continuous measurements that acquire quantum
information faster than the environment destroys it~\cite{PhysRevLett.100.013601,
PhysRevA.107.032410,RevModPhys.86.1391}.

Suspended mirrors offer an unusually favorable thermal-noise scale.
Mechanical loss is associated with elastic strain in the suspension, whereas
most of the pendulum-mode energy is stored in the conservative gravitational
potential, so gravitational dissipation dilution can reduce the modal loss far
below the intrinsic material loss~\cite{Cagnoli2000,10.1063/1.1148692}. The residual anelastic loss is well described by structural damping~\cite{Neben2012}. For an
approximately frequency-independent loss angle, the fluctuation--dissipation
theorem gives a thermal-force spectrum proportional to inverse frequency~\cite{Saulson1990,Gonzalez1994,Fedorov2018}. Nearly lossless trapping~\cite{Corbitt2007Cooling,Ni2012,PhysRevLett.122.071101,Whittle2021} can therefore shift the pendulum resonance upward and reduce the thermal noise sampled at the trapped resonance, but it does not whiten the bath: the same $1/f$ spectrum leaves an excess low-frequency tail.

This distinction is particularly important for entanglement.
Ground-state cooling primarily benefits from the reduced thermal-force scale near the
trapped resonance, whereas conditional entanglement depends on the transfer-function-weighted noise spectrum over a broader frequency range.
Colored noise has been included in measurement-based mechanical-state
preparation~\cite{Meng2022} and in optomechanical entanglement with non-Markovian
environments~\cite{Direkci2024}. The entanglement cost due to the off-resonant
color of structural noise, however, has not been isolated. Consequently,
the suspension improvement needed to offset the increase in the entanglement
threshold has remained unclear.

Here we isolate this cost and turn it into an experimental design target.
With $10\%$ detection loss, our model predicts a $28.9\%$ higher entanglement
threshold for structural noise than for a white bath with the same thermal-noise
level at the trapped resonance. We then realize a stepped fused-silica suspension with
an energy-decay rate $\Gamma/2\pi=361(39)$ nHz at the bare pendulum frequency $\omega_0/2\pi=2.63$ Hz. Its measured reduction in
$\omega_0\Gamma$ gives a suspension gain $G_q\simeq2.5$ relative to the previous
suspension (uniform reference)~\cite{PhysRevLett.124.221102}. This gain more than offsets the structural-noise penalty. With the covariance-reconstruction method of
Ref.~\cite{SakaiMatsumoto2026}, our multimode forecast gives a positive covariance-error allowance only for the stepped suspension: additional covariance error
can be tolerated while still establishing entanglement after band truncation
and experimental uncertainties are included.

\begin{figure*}[t!]
\centering
\includegraphics[width=0.9\textwidth]{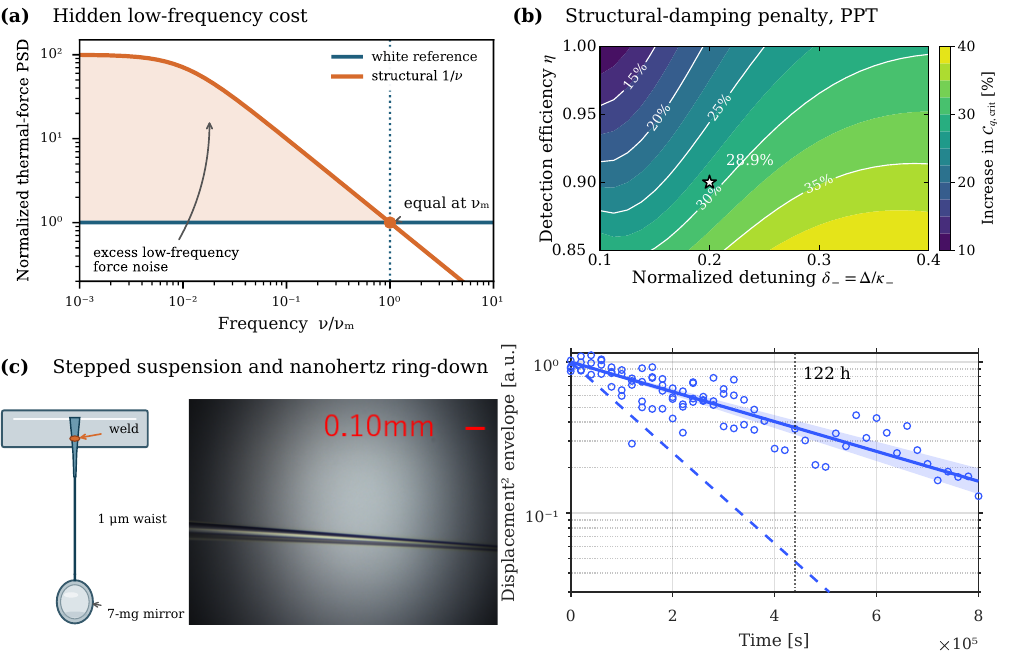}
\caption{Entanglement cost and measured suspension gain.
(a) White and regularized structural thermal-force spectra, normalized to their common value at the matching frequency $\nu_m$. The structural curve uses the illustrative cutoff ratio $\nu_c/\nu_m=0.01$.
(b) Structural-noise increase in critical cooperativity relative to the white bath matched at $\nu_m^\sigma$, versus normalized detuning $\delta_-$ and detection efficiency $\eta$, for linewidth ratio $\zeta=5$ and a circulating power of 6.39 W per arm, inferred from the ring-cavity experiment of Ref.~\cite{PhysRevLett.122.071101} with the revised calibration of Ref.~\cite{SakaiMatsumoto2026}. Star: $(\delta_-,\eta)=(0.2,0.9)$.
(c) Suspension schematic, fiber micrograph, and ring-down of the $7$-mg mirror. The ring-down signal is low-pass-filtered squared displacement in arbitrary units. Solid curve: joint fit; band: curves with $\Gamma$ shifted by $\pm$ one leave-one-record-out jackknife standard error. Dashed curve: previous suspension~\cite{PhysRevLett.124.221102}, with decay time 40.2 h. Dotted line: stepped decay time $1/\Gamma=122$ h.}
\label{fig:overview}
\end{figure*}

\textit{Entanglement penalty from structural damping.}---
We consider two identical suspended mirrors continuously measured in a
power-recycled Fabry--Perot Michelson interferometer with amplitude-quadrature
readout~\cite{PhysRevA.107.032410}. Their collective motion is decomposed into
the common mode ($\sigma=+$) and the differential
mode ($\sigma=-$).
To isolate the off-resonant spectral-shape penalty, we compare a structural bath with a white reference bath for each mode. We match their thermal-force spectra at the adiabatic optical-spring frequency $\nu_m^\sigma$ (the matching frequency), while holding the mechanical susceptibility and viscous damping fixed (End Matter, Single-mode comparison).
We use the same optical setting (circulating power, detuning, cavity linewidths, and detection efficiency) for both bath models.

We use symmetrized one-sided power spectral densities (PSDs) per hertz, with $\nu>0$ and $\omega=2\pi\nu$.
The thermal-force PSD $S_{FF,\sigma}^{\rm th}$ is constant for the white bath, whereas for the structural bath it is

\begin{equation}
 S_{FF,\sigma}^{\rm th}(\nu;\nu_c)=
 S_{FF,\sigma}^{\rm th}(\nu_m^\sigma)
 \frac{\sqrt{(\nu_m^\sigma)^2+\nu_c^2}}
      {\sqrt{\nu^2+\nu_c^2}},
 \label{eq:regularizedstructural}
\end{equation}
where the low-frequency cutoff is $\nu_c=1$ Hz. The spectrum approaches $1/\nu$ above $\nu_c$ and remains finite at zero frequency, keeping the low-frequency force variance finite. 
The structural bath has more force noise below $\nu_m^\sigma$
[Fig.~\ref{fig:overview}(a)] and less above it.

We evaluate the penalty with a closed-loop state-space model (End Matter, Dynamics). The structural
bath is represented by auxiliary Ornstein--Uhlenbeck states~\cite{UhlenbeckOrnstein1930}. To quantify the measurement strength relative to thermal noise, we define the quantum cooperativity of each collective mode as

\begin{equation}
 \mathcal C_q^\sigma(\nu)\equiv
 \frac{S_{FF,\sigma}^{\rm BA}(\nu)}
      {S_{FF,\sigma}^{\rm th}(\nu)}.
\end{equation}
Here $S_{FF,\sigma}^{\rm BA}$ is the back-action force PSD.
Following Ref.~\cite{PhysRevA.107.032410}, we use
$\mathcal C_q\equiv\mathcal C_q^-(\nu_m^-)$.
Both modes and their force--readout correlations enter the entanglement test.

The conditional steady-state mechanical covariances are transformed from
the collective modes to the individual-mirror basis, where entanglement is
evaluated by the Gaussian positive-partial-transpose (PPT) criterion~\cite{Simon2000}.
The PPT boundary is $\widetilde\mu_{\min}=1/2$, where $\widetilde\mu_{\min}$ is the
smallest symplectic eigenvalue of the partially transposed covariance
matrix and the vacuum covariance is $I/2$. The critical cooperativity $\mathcal C_{q,\rm crit}$
is the minimum $\mathcal C_q$ required to reach this boundary.

Figure~\ref{fig:overview}(b) shows the structural-noise increase in this
threshold relative to the matched white bath, as a function of normalized
differential detuning $\delta_-\equiv\Delta/\kappa_-$ and detection
efficiency $\eta$. Here $\Delta$ is the laser frequency minus the effective
cavity resonance, and $\kappa_\sigma$ is the full cavity linewidth, both in
angular-frequency units.
We set the power in each arm to the circulating power inferred for the ring cavity of Ref.~\cite{PhysRevLett.122.071101} using the revised calibration of Ref.~\cite{SakaiMatsumoto2026} (End Matter). For both collective modes ($\sigma=\pm$), this comparison retains only the pendulum mechanical mode. Power recycling narrows the common-mode linewidth~\cite{PhysRevA.107.032410}. We choose $\zeta\equiv\kappa_-/\kappa_+=5$ for this figure.
At the starred point in Fig.~\ref{fig:overview}(b), $(\delta_-,\eta)=(0.2,0.9)$, the structural bath
raises the threshold by $28.9\%$.

To offset this threshold increase, the required suspension gain is
\begin{equation}
 G_{\rm req}\equiv
 \frac{\mathcal C_{q,\rm crit}^{\rm struct}}
      {\mathcal C_{q,\rm crit}^{\rm white}}
 =1.289.
\end{equation}

To understand why the structural bath raises the entanglement threshold, we compare the conditional covariances for the two baths at fixed cooperativity. The structural bath increases the estimation-error variance primarily along the anti-squeezed quadrature of each collective mode. The conditional two-mirror state becomes less pure, and the two covariance ellipses rotate slightly in opposite directions. Together, these changes raise $\widetilde\mu_{\min}$ toward the PPT boundary. Because the optimal filter tracks much of the slow motion driven by excess force noise below the matching frequency, the lowest frequencies contribute little to the difference in $\widetilde\mu_{\min}$ between the two baths. At the starred point in Fig.~\ref{fig:overview}(b), frequencies up to the nominal 1-Hz cutoff account for less than 1\% of the full-band difference. In a direct test of the cutoff dependence, raising $\nu_c$ from 1 mHz to 10 Hz changes $G_{\rm req}$ only from 1.292 to 1.263, with the bath spectra matched at $\nu_m^\sigma$ throughout.

To express this requirement in terms of suspension parameters, we compare
suspensions at the same mirror mass $m$, temperature, and optical setting. These optical parameters fix the back-action
force spectrum. Because the optical spring dominates the bare pendulum
stiffness, the optical-spring frequencies are nearly the same for the two
suspensions. For structural damping,
$S_{FF}^{\rm th}(\nu_m^\sigma)\propto m\omega_0\Gamma/\nu_m^\sigma$.
The stepped suspension therefore changes $\mathcal C_q$ by a factor equal to the
suspension gain
$G_q=(\omega_0\Gamma)_{\rm uniform}/(\omega_0\Gamma)_{\rm step}$
(step: stepped suspension; uniform: previous suspension modeled with a uniform
fiber). If $G_q>G_{\rm req}$, this gain exceeds the threshold increase caused
by structural noise.

\textit{Stepped-fiber fabrication}.---Thin fibers enhance gravitational dissipation dilution, but their larger surface-to-volume ratio increases surface loss~\cite{10.1063/1.1148692,Gretarsson1999,PENN20063}. To achieve strong dissipation dilution
while mitigating the additional surface loss introduced by laser welding,
we fabricate a stepped fused-silica suspension fiber using the flame-brush
tapering technique of Ref.~\cite{Nagai:14}.

\begin{figure}[t!]
\centering
\includegraphics[width=\columnwidth]{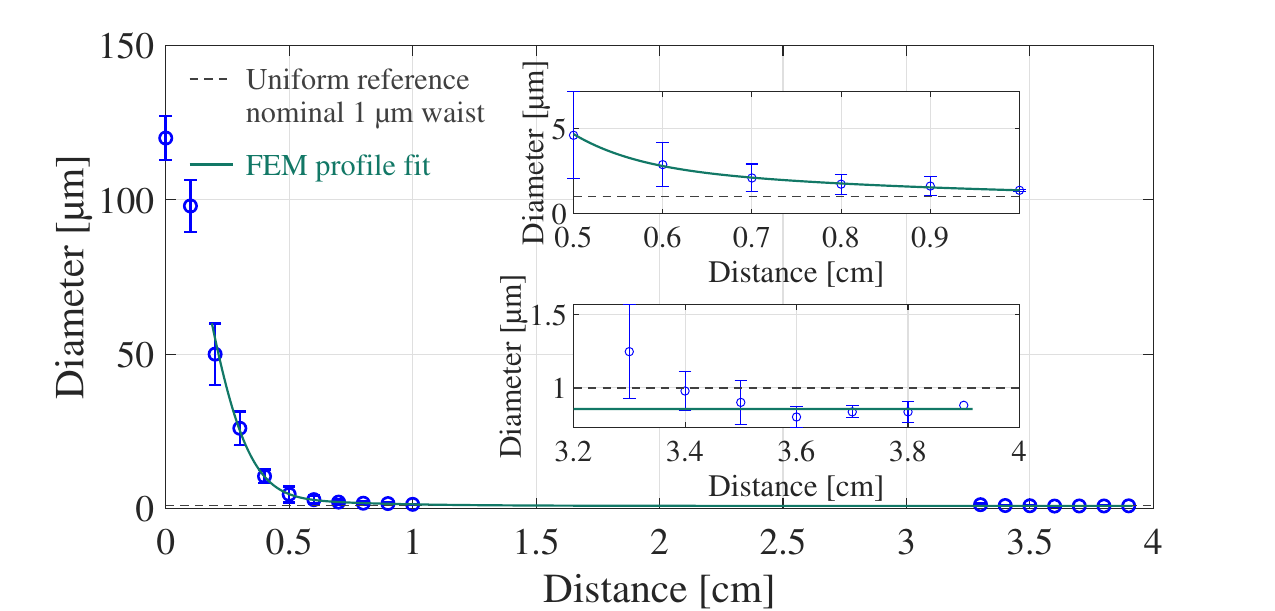}
\caption{Diameter profile of the stepped fiber. Blue points show measurements on five fibers; vertical bars indicate standard deviations. Insets enlarge the waist and upper section. Solid curve: fitted profile used in the finite-element model (FEM). Dashed lines: the previous suspension's specified $1\,\mu$m waist diameter~\cite{PhysRevLett.124.221102}.}
\label{fig:fiber-profile}
\end{figure}

Starting with a $125\,\mu$m-diameter silica fiber, we heat the fiber with
a scanning flame and pull it from both ends using motorized translation
stages. By programming the scan-length profile and stopping the taper at the
target waist, we produce a waist with a target diameter of $1\,\mu$m and
length of $3.5$ cm, connected to a nearly unmodified thick upper section.
Profile measurements of five fibers are shown in Fig.~\ref{fig:fiber-profile}.

To form the pendulum, a thick section of the stepped fiber, approximately
$60\,\mu$m in diameter, is welded to a silica plate, while the thin section
is epoxy-bonded to the silica mirror.

\needspace{3\baselineskip}
The previous suspension was
welded at both ends~\cite{PhysRevLett.124.221102}. Because the attachment
methods also differ, the measured gain compares the complete suspensions.
In the modeled stepped geometry, little fundamental-mode bending energy
is stored near the epoxy-bonded lower attachment; bending in the taper and upper section
is retained in the loss calculation.

\textit{Nanohertz ring-down.}---The fundamental pendulum mode is excited
and allowed to decay freely in high vacuum ($\sim10^{-5}$ Pa), while its
displacement is monitored interferometrically. 
Four ring-down records, totaling approximately $516$ h after removal of filter-edge transients, are fitted jointly [Fig.~\ref{fig:overview}(c)]. Their low-pass-filtered squared signals are fitted to exponential decays with a common energy-decay rate $\Gamma$ and record-dependent amplitudes. At $\omega_0/2\pi=2.63$ Hz, we measure $\Gamma/2\pi=361(39)$ nHz, corresponding to $Q_0\equiv\omega_0/\Gamma=7.3(8)\times10^6$. Parentheses give the leave-one-record-out jackknife standard error.

The measured suspension parameters give $G_q=2.5(3)$, about twice the required gain. The uncertainty combines the independent decay-rate uncertainties of the two suspensions, with their frequencies held fixed.
At the starred point in Fig.~\ref{fig:overview}(b), the single-mode model gives
$\mathcal C_q\simeq18.2$ and $46.5$ for the uniform and stepped suspensions,
respectively. Both exceed the structural-bath threshold
$\mathcal C_{q,\rm crit}=2.305$. Thus, the model predicts conditional entanglement
for both suspensions when verification errors are neglected.

\Needspace{3\baselineskip}
\textit{Residual loss}.---For the fitted stepped profile, the finite-element model (FEM) assigns $0.79\%$ of the fundamental mode's total potential energy to elastic bending. The dilution factor is the inverse of this fraction, giving about 126. Integrating the surface, bulk, and thermoelastic loss kernels (End Matter, Geometry and loss) over this profile gives a material-limited quality factor of $3.4\times10^7$ before adding the fitted attachment loss. This material-loss prediction is about $4.6$ times the measured $Q_0$, indicating residual loss beyond the adopted material model.

A plausible source is the upper attachment: laser welding from above visibly deformed the thick section, increasing its surface-to-volume ratio and potentially concentrating bending strain. Device-to-device reproducibility of $Q_0$ has not yet been assessed; tests on multiple devices with lateral upper welds that preserve the cross section are planned to examine both this loss attribution and reproducibility.

\Needspace{3\baselineskip}
\textit{Feasibility of entanglement verification}.---Generating entanglement is not enough: it must also be established from a band-limited record of finite length with imperfect calibration. We therefore ask how much error the reconstructed covariance can carry before the entanglement test fails, and compare this tolerance with the errors expected for the uniform reference~\cite{PhysRevLett.124.221102} and the stepped suspension. We use the three-trajectory method of Ref.~\cite{SakaiMatsumoto2026},
an unbiased conditional-covariance estimator. We integrate the one-sided PSD of the difference between the forward and smoothed trajectories over 130--1990 Hz. The forward trajectory uses the past record, and the smoothed trajectory also uses the future record. We lower the upper edge of the 130--2000 Hz band used in Ref.~\cite{SakaiMatsumoto2026} to 1990 Hz to exclude the uniform suspension's first violin resonance at 1994 Hz. The stepped suspension's first violin resonance is at 2936 Hz.

For the forecast, we require the covariance estimate predicted from the finite analysis band to satisfy the quantum uncertainty relation (End Matter, Verification). Entanglement is established only if every covariance within the error bounds that satisfies this relation is entangled. We summarize the outcome by the covariance-error allowance: the error radius that this test permits minus the radius consumed by the error budget, both expressed as fractions of the variance predicted without a band restriction, in every quadrature. A positive allowance is margin left for effects outside the model; a negative allowance means that the listed errors alone exceed what the test tolerates.

\begin{table*}[t]
\caption{Covariance-error allowance as effects are included one at a time, for optical settings optimized separately for each suspension. Rows 1--3: physical model; row 4: finite analysis band; rows 5--7: experimental uncertainties. Row 5 gives the Gaussian distributions of the common and differential displacement scales $a_{\rm cal,\pm}$; row 6 fixes the record duration $T$. Row 7 gives standard deviations of zero-mean Gaussian parameter errors (relative except for $\delta_-$; uniform/stepped where two values are given). Changes are in percentage points (pp); entries are rounded independently. Bold: final allowance. $\Gamma_0$ is the bare energy-decay rate. Conventions and coverage are specified in the End Matter.}
\label{tab:certification_budget}
\begin{ruledtabular}
\begin{tabular}{lrrrr}
 &\multicolumn{2}{c}{Uniform}&\multicolumn{2}{c}{Stepped}\\
Effect included & Change (pp) & Allowance (\%) & Change (pp) & Allowance (\%)\\
\hline
White bath, single mode (starting allowance) &--- &$90.8$ &--- &$105.8$\\
Spectral color &$-16.2$ &$74.6$ &$-10.7$ &$95.1$\\
Pitch, internal modes, and target change &$-41.8$ &$32.8$ &$-14.8$ &$80.3$\\
Band truncation, 130--1990 Hz &$-9.9$ &$22.9$ &$-8.4$ &$71.9$\\
Displacement-scale calibration, $10.5\%$: $a_{\rm cal,\pm}\sim\mathcal N(1,0.105^2)$~\cite{SakaiMatsumoto2026} &$-31.1$ &$-8.2$ &$-42.3$ &$29.6$\\
Finite-record statistics, $T=1.64$ s~\cite{SakaiMatsumoto2026} &$-2.2$ &$-10.4$ &$-3.2$ &$26.4$\\
Parameters: $\eta$ $0.4\%$, $\delta_-$ $0.0017$, $\nu_m^-$ $0.5\%$, $\Gamma_0$ $4.0/10.8\%$ &$-0.2$ &$\mathbf{-10.7}$ &$-1.6$ &$\mathbf{24.8}$\\
\end{tabular}
\end{ruledtabular}
\end{table*}

Table~\ref{tab:certification_budget} builds the allowance in three stages. Rows 1--3 make the physical model realistic, starting from a single pendulum mode in a white bath (row 1). Unlike the idealized $1/\nu$ bath of Fig.~\ref{fig:overview}(b), the structural bath in row 2 includes surface, bulk, and thermoelastic losses~\cite{PENN20063,Cumming2020Loss,Bell2014Thermoelastic}, together with attachment loss fitted to each suspension's measured fundamental decay. Row 3 adds pitch and internal (violin) modes of the fiber from a beam FEM using the same loss distribution, including optical-spring coupling between pendulum and pitch~\cite{PhysRevA.107.033515} and back-action-induced thermal-noise correlations~\cite{Ma2026Broadband}. It also changes the covariance target from pendulum motion to the combined pendulum--pitch motion. Row 3 reduces the allowance by 41.8 percentage points for the uniform suspension and 14.8 for the stepped suspension. Removing the first violin mode from the full model restores 38.6 and 11.3 points, respectively.

Row 4 centers the error bounds on the covariance predicted from the analysis band and includes its full difference from the full-band covariance. Rows 5--7 add calibration, finite-record, and parameter uncertainties using the inputs in Table~\ref{tab:certification_budget}. The calibration uncertainty and record duration are taken from Ref.~\cite{SakaiMatsumoto2026}; the other inputs are an assumed $0.4\%$ relative standard uncertainty in total detection efficiency, guided by the photodiode-responsivity calibration of Ref.~\cite{Shaw2000}, a detuning uncertainty from the spread observed in Ref.~\cite{SakaiMatsumoto2026}, an assumed $0.5\%$ optical-spring-frequency uncertainty, and the measured decay-rate uncertainties. Among these uncertainty contributions, calibration gives the largest reduction in allowance.

At the circulating power used in Fig.~\ref{fig:overview}(b), we selected the optical settings separately for each suspension over positive detunings and linewidth ratios $1\leq\zeta\leq42$. In a lossless, mode-matched model~\cite{PhysRevA.107.032410}, the measured recycling-mirror power reflectivity of $0.91(1)$~\cite{Pradyumna2020} gives $\zeta\simeq42.4$, motivating the cap of 42 reached by the stepped suspension. Both collective optical-spring frequencies lie inside the analysis band for each suspension (574 Hz and 1.74 kHz for the stepped one).

\needspace{6\baselineskip}
The stepped suspension retains a 24.8\% allowance,
whereas the uniform reference has a 10.7\% deficit. At these settings, with other uncertainties fixed, the allowance vanishes at relative standard uncertainties of 6.4\% (uniform) and 19.4\% (stepped) in displacement scale. The corresponding limit for detection efficiency is 3.8\% for the stepped suspension (End Matter, Conventions of Table~\ref{tab:certification_budget}). The uniform deficit remains 9.3\% for an arbitrarily long record, with the other uncertainties fixed. Thus, in this forecast, the $10.5\%$ calibration uncertainty achieved in Ref.~\cite{SakaiMatsumoto2026} suffices for the stepped suspension but not for the uniform reference.

Finally, at the starred point in Fig.~\ref{fig:overview}(b) ($\zeta=5$, $\delta_-=0.2$), the nominal covariance reconstructed from the 130--1990 Hz band for the stepped suspension does not satisfy the quantum uncertainty relation. At this lower recycling gain, widening the analysis band to 100--5100 Hz, for example, yields a physical covariance and leaves an 8.7\% covariance-error allowance under the full error budget of Table~\ref{tab:certification_budget}.

\textit{Summary}.---Structural damping imposes a low-frequency entanglement
cost invisible to a resonance-only comparison of thermal-force noise.
The model predicts a $28.9\%$ increase in critical cooperativity at the starred point in Fig.~\ref{fig:overview}(b)
($G_{\rm req}=1.289$). Our stepped fused-silica suspension achieves
$\Gamma/2\pi=361(39)$ nHz, giving $G_q\simeq2.5$, about twice this requirement.
Using the experimental calibration and record-length benchmarks of Ref.~\cite{SakaiMatsumoto2026} and the stated parameter uncertainties, we optimize the optical settings separately for each suspension under the quantum uncertainty relation. With band truncation included, the multimode forecast leaves only the stepped suspension a positive covariance-error allowance of $24.8\%$. This allowance remains available for additional model errors and technical noise.

{\it Acknowledgments.---}
We thank Hirotomo Karino for early contributions, and Katsuta Sakai, Akihisa Goban, Takao Aoki, and Mark Sadgrove for discussions. N.M. is supported by JST FOREST Grant No.~JPMJFR202X.
We used GPT-6 Astra (OpenAI) and Claude Fable 5.1 and 5.2 (Anthropic)
to assist with analysis and manuscript editing.
The authors verified the results, reviewed the AI-assisted text,
and take responsibility for the final content.

\paragraph*{Data availability.---}
The four ring-down measurement records are openly available on Zenodo~\cite{Sawada2026Ringdown}.
The code, material constants, finite-element inputs, and numerical results for Tables~I and II are in the Zenodo computational archive (\href{https://doi.org/10.5281/zenodo.24442034}{doi:10.5281/zenodo.24442034}).

\makeatletter
\auto@bib@empty
\begingroup
\let\auto@bib@innerbib\@empty
\makeatother

\endgroup

\onecolumngrid
\clearpage
\twocolumngrid
\appendix
\providecommand{\sym}{\operatorname{sym}}
\section{End Matter}
\label{app:calculation}
\begingroup
\clubpenalty=10000\widowpenalty=10000

The first three parts specify the model behind Fig.~\ref{fig:overview}(b) and rows 1--3 of Table~\ref{tab:certification_budget}; the last three define the verification test behind rows 4--7. Complete expressions, numerical settings, and code are in the computational archive (see Data availability).

\textit{Dynamics}.---To calculate the conditional covariances, we extend the model of Ref.~\cite{PhysRevA.107.032410} to multiple mechanical modes and colored thermal noise. For the common and differential channels ($\sigma=\pm$), with mechanical quadratures $q_j,p_j$, cavity amplitude/phase quadratures $X,Y$, and readout $y_\sigma$,
\begin{align}
 \dot q_j&=\omega_jp_j,\nonumber\\
 \dot p_j&=-\omega_jq_j-\delta_{j0}k_cq_0/\omega_0-2g_jX\nonumber\\
 &\quad +(g_j/g_0)f_{{\rm fb},\sigma}+\sum_aF_{ja}z_a,\nonumber\\
 \dot X&=-\kappa_\sigma X/2-\Delta Y+\sqrt{\kappa_\sigma}x^{\rm in},\nonumber\\
 \dot Y&=-\kappa_\sigma Y/2+\Delta X-2\sum_jg_jq_j
       +\sqrt{\kappa_\sigma}y^{\rm in},\nonumber\\
 y_\sigma&=-\sqrt{\eta\kappa_\sigma}X+
 \sqrt\eta\,x^{\rm in}+\sqrt{1-\eta}\,v.
 \label{eq:components}
\end{align}
Mode $j$ has bare frequency $\omega_j$ ($j=0$: pendulum) and coupling $g_j$. We use $[q_i,p_j]=\mathrm{i}\delta_{ij}$ and $[X,Y]=\mathrm{i}$. The independent vacuum inputs $x^{\rm in},y^{\rm in},v$ (amplitude, phase, loss) have white-noise intensity $1/2$ and unit one-sided PSD. Auxiliary Ornstein--Uhlenbeck states $z_a$ represent the thermal bath and couple through $F_{ja}$; the stiffness counterterm $k_c$ fixes the free pendulum resonance.

Calculations for both suspensions use bath temperature $T_b=300$ K, $\eta=0.90$, and $\kappa_-/2\pi=1.640$ MHz. For the ring cavity of Ref.~\cite{PhysRevLett.122.071101}, the revised calibration of Ref.~\cite{SakaiMatsumoto2026} gives a mean intracavity photon number $\bar n=1.1273\times10^{10}$ and round-trip length $L_{\rm rt}=0.09881$ m. At $\lambda=1064$ nm, the circulating power is $\bar n(hc/\lambda)c/L_{\rm rt}=6.38549$ W. We use this power in each normal-incidence, 0.10-m model arm. Table~\ref{tab:material_parameters} gives the corresponding coupling $g_{\rm ref}$ at normalization frequency $\omega_{\rm ref}$. Selected settings are $(\zeta,\delta_-)=(16.3,0.156439)$ (uniform) and $(42,0.174320)$ (stepped).

A causal lead--lag controller (specified in the archive) uses $y_\sigma$ to generate the stabilizing drive $f_{{\rm fb},\sigma}$. Because the estimator contains the same dynamics and no independent actuator noise is added, this feedback changes neither the conditional covariance nor the forward-minus-smoothed trajectory difference.

\begin{table}[b!]
\caption{Optical inputs and suspension fits (uniform/stepped; extra digits specify inputs). Fused-silica constants are listed in the computational archive.}
\label{tab:material_parameters}
\begin{ruledtabular}
\begin{tabular}{ll}

Power per arm & $6.38549$ W\\
$g_{\rm ref}/2\pi$ & $287.356$ kHz\\
$\omega_{\rm ref}/2\pi$ & $2.63$ Hz\\
$\Gamma_0/2\pi$ & $1103.60/361$ nHz\\
Fitted $\phi_A$ & $0.00775356/0.00203250$\\
$k_c$ ($10^{-5}$ s$^{-2}$) & $-9.38302$ / $-4.04400$
\end{tabular}
\end{ruledtabular}
\begingroup\raggedright
Here $\Gamma_0=\omega_0/Q_0$ ($\Gamma$ for stepped): the uniform value uses the FEM frequency and measured $Q_0=2.0\times10^6$; the stepped value is measured by ring-down.

\par
\endgroup
\end{table}

\textit{Geometry and loss}.---Each fiber is modeled as a prestressed Euler--Bernoulli beam with self-weight and a fixed upper weld. The uniform fiber is 50 mm long and 1 $\mu$m in diameter; the stepped fiber follows the fitted profile of Fig.~\ref{fig:fiber-profile}. The 14 mode shapes $\Phi_j$ (pendulum, pitch, 12 violin) satisfy $\Phi_i^T M\Phi_j=m\delta_{ij}$, where $M$ is the FEM mass matrix. Their overlaps $b_j=\Phi_j(\ell_f)+h_m\Phi'_j(\ell_f)\geq0$ give the couplings $g_j=g_{\rm ref}b_j\sqrt{\omega_{\rm ref}/\omega_j}$. Here $\ell_f$ is the total modeled fiber length and $h_m$ the mirror's center-of-mass offset below attachment. FEM fundamental/pitch frequencies are 2.21/27.74 Hz (uniform) and 2.63/27.83 Hz (stepped).

\Needspace{4\baselineskip}
Loss in each suspension fiber produces correlated thermal forces on its different modes. With $\omega_c=2\pi\nu_c$, the correlated canonical-force spectrum is
\begin{equation}
 [\mathcal S_p(\nu)]_{ij}
 =\frac{4k_BT_b}{\hbar\sqrt{\omega_i\omega_j}}
 \left[\sum_e\mathsf B^e_{ij}\mathcal K_e(\omega)
 +\frac{\phi_A\mathsf B^A_{ij}}{\sqrt{\omega^2+\omega_c^2}}\right].
 \label{eq:multimode_force}
\end{equation}
The first term is the material loss. For element diameter $d_e$ and Young's modulus $E$, the bending matrix is $\mathsf B^e_{ij}=m^{-1}\int_e E(\pi d_e^4/64)\Phi_i''\Phi_j''dx$, and the element kernel $\mathcal K_e(\omega)$ sums surface, bulk, and thermoelastic losses~\cite{Gretarsson1999,PENN20063,Cumming2020Loss,Bell2014Thermoelastic} (explicit kernels in the archive). The second term is the attachment loss: the matrix $\mathsf B^A_{ij}$ restricts the bending integral to the upper 200 $\mu$m, with fitted excess loss angle $\phi_A\geq0$. The surface and bulk terms are regularized at $\nu_c$ like the attachment term.

The states $z_a$ approximate this spectrum with memory friction consistent with the fluctuation--dissipation theorem. We fit $\phi_A$ and $k_c$ to the bare pendulum frequency and energy-decay rate. Rows 1--2 of Table~\ref{tab:certification_budget} retain only pendulum dynamics; rows 3--7 retain all 14 modes. Predicted quality factors are $(5.2,6.2)\times10^6$ for pitch and $(8.9,9.5)\times10^5$ for the first violin mode (uniform/stepped). The white bath matches the thermal spectrum at $\nu_m^\sigma$, retaining the mechanical response. At the optical settings in Table~\ref{tab:certification_budget}, lowering $\nu_c$ from 1 Hz to 1 mHz and refitting to preserve the bare pendulum frequency and decay rate raises the row-2 allowance by 2.15 and 0.80 percentage points for the uniform and stepped suspensions, respectively.

\textit{Single-mode comparison}.---Figure~\ref{fig:overview}(b) retains only the pendulum for each $\sigma=\pm$, with damping $-\Gamma_0p_0$ and coupling $g_0=g_{\rm ref}\sqrt{\omega_{\rm ref}/\omega_0}$ ($b_0=1$ in this single-mode approximation). At the adiabatic optical-spring frequency $(\omega_m^\sigma)^2=\omega_0^2+4\omega_0g_0^2\Delta/[(\kappa_\sigma/2)^2+\Delta^2]$, the matched canonical PSD is $S_{FF,\sigma}^{\rm th}(\nu_m^\sigma)/(\hbar m\omega_0)=4k_BT_b\Gamma_0/(\hbar\omega_m^\sigma)$. At the starred point in Fig.~\ref{fig:overview}(b), the stepped optical-spring frequencies are $\nu_m^-=604$ Hz and $\nu_m^+=1.456$ kHz. Because $\mathcal C_q^+(\nu_m^+)/\mathcal C_q^-(\nu_m^-)\simeq2.79$, we quote the smaller, differential-mode value as $\mathcal C_q$. For the uniform reference, this comparison uses the measured $\omega_0/2\pi=2.2$ Hz and $Q_0=2.0\times10^6$ of Ref.~\cite{PhysRevLett.124.221102}.

\textit{Reconstruction}.---The entanglement test uses one canonical pair per mirror, built from the center-of-mass contribution of a set of modes $L$: pendulum alone ($L=\{0\}$) in rows 1--2, and pendulum plus pitch ($j=1$; $L=\{0,1\}$) in rows 3--7. In each collective channel, with $\sigma$ suppressed, the overlaps $b_j$ give
\begin{equation}
 \begin{pmatrix}q_L\\p_L\end{pmatrix}
 =\frac{\sum_{j\in L}b_j
 \begin{pmatrix}\sqrt{\omega_0/\omega_j}q_j\\\sqrt{\omega_j/\omega_0}p_j\end{pmatrix}}
 {\sqrt{\sum_{j\in L}b_j^2}},\quad[q_L,p_L]=\mathrm{i}.
 \label{eq:low_translation}
\end{equation}
The two-mirror coordinates are the sum and difference of the common and differential coordinates, divided by $\sqrt2$.

We use the past-conditioned center and finite-band covariance of Ref.~\cite{SakaiMatsumoto2026}, Eqs.~(12) and (21), respectively. A hat denotes a record estimate, $B=[130,1990]$ Hz the analysis band, and subscript $0$ the stated inputs before uncertainty propagation.

With the trajectory filters and reconstruction gain fixed, the covariance estimate depends on the record through $\widehat D_B$, the sample covariance over $B$ of the forward-minus-smoothed trajectory difference $\bm a$ for the two mirrors. Here $S_{\bm a}$ is the PSD matrix of $\bm a$ and $D_B$ is the model mean of $\widehat D_B$. For a stationary model with filters matched to its dynamics and noise, $D=\operatorname{Cov}(\bm a)\succ0$, and $V$ is the conditional covariance given the past record. We compute the estimate $\widehat V$ with the nominal gain $D_0^{-1}V_0$ and the finite-band prediction $P_B$ with model-matched filters and gain.
\begin{align}
 D_B&=\operatorname{Re}\!\int_B S_{\bm a}(\nu)\,d\nu,\nonumber\\
 \widehat V&=\sym(\widehat D_BD_0^{-1}V_0),\nonumber\\
 P_B&=\sym(D_BD^{-1}V),
 \label{eq:finite_band_map}
\end{align}
where $\sym A=(A+A^T)/2$. Under the statistical model below, the mean reconstruction at the stated inputs is $P_{B,0}$. Using the full spectrum with matched filters and gain would give $P_B=V$. Restricting it to $B$ leaves the band-truncation difference $\Delta_{B,0}=V_0-P_{B,0}$, which enters the error budget.

\looseness=-1 \textit{Verification}.---The data determine the covariance only to within an error interval; entanglement is established if even its least favorable member is entangled. Consider possible covariance matrices $W$ within $\widehat V-rV_0\preceq W\preceq\widehat V+rV_0$, where $r\geq0$ is the dimensionless error radius relative to $V_0\succ0$.  Assume the interval contains at least one physical covariance, $W+\mathrm{i}\Omega/2\succeq0$, where $\Omega=\bigoplus_{k=1}^2\left(\begin{smallmatrix}0&1\\-1&0\end{smallmatrix}\right)$. The upper endpoint is then physical and gives the most stringent PPT test~\cite{BhatiaJain2015}. Every Gaussian state in the interval is therefore entangled exactly when
\begin{equation}
 \widetilde\mu_{\min}(\widehat V+rV_0)<\tfrac12.
 \label{eq:cert_condition}
\end{equation}
The forecast sets $\widehat V=P_{B,0}$. We use $1/2+0.0002$ as the search threshold for the smallest symplectic eigenvalue of $P_{B,0}$ before partial transposition. At both selected settings, this eigenvalue lies just above the threshold.

The allowance compares two radii, measured in the relative norm $\|A\|_{V_0}=\|V_0^{-1/2}AV_0^{-1/2}\|_2$, where $\|\cdot\|_2$ is the operator norm. The first, $r_{\rm err}$, is the radius consumed by three error matrices, with signs chosen so that their sum is the true-minus-reconstructed covariance: (i) the band-truncation difference $\Delta_{B,0}$, with $\epsilon_B=\|\Delta_{B,0}\|_{V_0}=0.575/0.427$ (uniform/stepped); (ii) $\Delta_{\rm stat}$, from finite-record fluctuations of $\widehat D_B$ at the stated inputs, approximated as jointly Gaussian with covariance inversely proportional to the record duration $T$; and (iii) $\Delta_{\rm par}$, the change from $\Delta_{B,0}$ caused by the parameter uncertainties of Table~\ref{tab:certification_budget}, evaluated with the controller, trajectory filters, and reconstruction gain held fixed. The three matrices are combined before taking the norm:
\begin{align}
 r_{\rm err}=\max\bigl\{&\epsilon_B,
 \label{eq:full_error_radius}\\
 &\operatorname{Quant}_{0.975}\|\Delta_{B,0}+\Delta_{\rm par}+\Delta_{\rm stat}\|_{V_0}\bigr\},\nonumber
\end{align}
where $\operatorname{Quant}_{0.975}$ is the 97.5th percentile and the lower bound $\epsilon_B$ retains the full band-truncation difference.

The second, $r_{\max}$, is the radius that the entanglement test permits once calibration uncertainty is included. Scaling the common and differential channels by $a_{\rm cal,+},a_{\rm cal,-}$ multiplies $\widetilde\mu_{\min}$ exactly by $|a_{\rm cal,+}a_{\rm cal,-}|$. For independent Gaussian scales with mean 1 and standard deviation 0.105 (Table~\ref{tab:certification_budget}, row 5), $c_{\rm cal}=\operatorname{Quant}_{0.975}(|a_{\rm cal,+}a_{\rm cal,-}|)=1.307$. The permitted radius $r_{\max}$ satisfies
\begin{equation}
 c_{\rm cal}\widetilde\mu_{\min}(P_{B,0}+r_{\max}V_0)=\tfrac12.
 \label{eq:residual_allowance}
\end{equation}
The allowance reported in Table~\ref{tab:certification_budget} is the remainder $r_{\rm res}=r_{\max}-r_{\rm err}$. The two 97.5\% bounds give at least 95\% joint coverage under the adopted uncertainty model.

\textit{Conventions of Table~\ref{tab:certification_budget}.}---Rows 1--3 use $[2\widetilde\mu_{\min}(V_0)]^{-1}-1$; row 4 sets $c_{\rm cal}=1$ and centers the permitted radius on $P_{B,0}$ instead of $V_0$, giving $r_{\max}-\epsilon_B$ (stepped: $114.6\%-42.7\%=71.9\%$).

The independent calibration and physical inputs are fixed within each record, treated as continuous and stationary. The stated standard deviation for $\eta$ precedes truncation to $[0,1]$. The temporal detuning spread in Sec.~S1.1 of the Supplemental Material of Ref.~\cite{SakaiMatsumoto2026} is $0.0034$ in half-linewidth units, or $0.0017$ in our full-linewidth units. With calibration and estimator settings fixed, we simulated the waveform-frequency estimator of Sec.~S2.1 of that supplement. Its maximum relative root-mean-square error of $0.49\%$ motivates the assumed $0.5\%$ frequency uncertainty.

At fixed optics, the calibration and efficiency standard-deviation limits solve $r_{\rm res}=0$. The long-record limit sets $\Delta_{\rm stat}=0$, with other uncertainties unchanged.

\endgroup

\end{document}